\documentclass[reprint,amsmath,amssymb,notitlepage,tightenlines,pra,unsortedaddress]{revtex4-1}
\usepackage{bm}
\usepackage{amsmath}
\usepackage{algorithm}
\usepackage{amssymb}
\usepackage{amsfonts}
\usepackage{physics}
\usepackage{graphicx}
\usepackage[noend]{algpseudocode}
\usepackage[large]{subfigure}
\usepackage{float}
\usepackage{caption}
\usepackage{color}

\usepackage{cleveref}

\newcommand{\rev}[1]{{\color{black} #1}}

\definecolor{violet}{rgb}{1,0,1}

\newcommand{\lela}{\left\langle}
\newcommand{\rira}{\right\rangle}
\graphicspath{{./energycomp}}
\graphicspath{{./excitecomp}}
\graphicspath{{./energydiff}}

\begin{document}

\title{Simulation of a hydrogen atom in laser field using the
time-dependent variational principle}

\author{Keefer Rowan*}
\author{Louis Schatzki*}
\author{Timothy Zaklama*}
\affiliation{Department of Physics and Astronomy, Vanderbilt University, Nashville, Tennessee, 37235, USA}

\author{*These authors were of equal contribution}
\affiliation{}

\author{Yasumitsu Suzuki}
\affiliation{Department of Physics, Tokyo University of Science,
1-3 Kagurazaka, Shinjuku, Tokyo 162-8601, Japan}

\author{Kazuyuki Watanabe}
\affiliation{Department of Physics, Tokyo University of Science,
1-3 Kagurazaka, Shinjuku, Tokyo 162-8601, Japan}

\author{K\'alm\'an Varga}
\affiliation{Department of Physics and Astronomy, Vanderbilt University, Nashville, Tennessee, 37235, USA}

\begin{abstract}
The time-dependent variational principle is used to optimize the
linear and nonlinear parameters of Gaussian basis functions to
solve the time-dependent Schr\"odinger equation in 1 and 3 dimensions 
for a one-body soft Coulomb potential in a laser field. 
The accuracy is tested comparing the solution to finite difference grid calculations 
using several examples. \rev{The approach is not limited to one particle
systems and the example presented for two electrons demonstrates the
potential to tackle larger systems using correlated basis functions.}
\end{abstract}

\maketitle

\section{Introduction}
Three dimensional or radial grids \cite{PhysRevA.88.023422} are flexible representations of
time-dependent wave functions in solutions of time-dependent problems
or in time-dependent density functional 
calculations. Localized basis functions, e.g. Gaussian orbitals, are 
less flexible for time-dependent Hamiltonians, 
e.g. for systems interacting with  strong laser pulses. In describing
ionization, one often needs to represent the wave function up to a few
hundred Bohr distances, requiring large spatial grids. The
attractive feature of basis functions in comparison to real space grids is reduced dimensionality. 
The question is how can we optimize the basis
functions to represent the rapidly changing  time-dependent wave
function. 

Due to the experimental advances in attosecond extreme ultraviolet light
pulses and intense x-ray sources \cite{RevModPhys.81.163},
many different basis function representations have been
developed to solve the time-dependent Schr\"odinger-equation for atoms
interacting with strong laser pulses
\cite{PhysRevA.90.033403,PhysRevA.90.012506,doi:10.1063/1.2358351,
PhysRevA.61.053411,PhysRevA.98.023413,PhysRevA.89.033415,
PhysRevA.77.033412,PhysRevA.65.063403}. The most often used basis functions are the
discrete variable representations \cite{PhysRevA.55.3417,
PhysRevLett.98.073001,PhysRevA.55.3417,PhysRevA.79.012719} and
B-splines \cite{Bachau_2001,PhysRevA.74.052702,PhysRevLett.103.063002}. 
These basis functions have been combined with innovative approaches to
solve the time-dependent Schr\"odinger equation
\cite{PhysRevE.90.063309,PhysRevA.60.3125,PhysRevA.95.023401,PhysRevA.78.032502,
doi:10.1063/1.466058,GHARIBNEJAD2019,PhysRevE.95.053309,Wells2019,kormann_2016,
doi:10.1063/1.465362,PhysRevA.99.013404,PhysRevE.95.023310}.
In these works, the  proper boundary conditions are enforced by using complex
absorbing potentials \cite{PhysRevA.78.032502,Yu_2018,DeGiovannini2015},
exterior complex scaling \cite{PhysRevA.81.053845,WEINMULLER2017199},
or perfectly matched layers \cite{SCRINZI201498}.

Gaussians basis functions are the most popular choices of quantum
mechanical calculations because their matrix elements can be evaluated
analytically \cite{RevModPhys.85.693,svm_book}.  
Gaussian functions however, have difficulties
in reproducing the characteristic oscillatory behavior
of continuum orbitals in the asymptotic region. 
Gaussians with complex
parameters may be better suited to describe the continuum because of their
inherent oscillatory nature \cite{PhysRevA.99.012504}. One way to
extend Gaussians for problems involving ionization  is to augment them
with suitable functions such as B-splines \cite{PhysRevA.90.012506}.

In this work we will solve the time-dependent Schr\"odinger-equation
by time propagation using a time-dependent basis. The parameters of
the basis will be optimized using the time-dependent variational
principle (TDVP) \cite{dirac_1930,doi:10.1080/00268976400100041}. 
We will consider a hydrogen atom in laser a field. The oscillating
field moves the electron density away from the atom and then back
towards the atom. The time-dependent basis functions will  be optimized to accurately
represent the moving density. 
 
The time-dependent variational method was  introduced by Dirac
\cite{dirac_1930}, extended by McLachlan
\cite{doi:10.1080/00268976400100041}
and reformulated for Gaussian wave packets in Ref.
\cite{doi:10.1063/1.449204}.
The time-dependent variational method has been used in various
calculations, such as in the description of the dynamical behavior of  Bose-Einstein condensates 
\cite{PhysRevA.82.023611}, and in wave packet
dynamics \cite{ZOPPE2005308,PhysRevA.79.043417}. Furthermore, the study of the dynamics of strongly interacting lattice bosons
\cite{Carleo2012} and strongly correlated electrons \cite{PhysRevB.92.245106} reflect the increasing popularity of the time dependent variational method in other fields.

The TDVP is also  often used in approximating complex many-body wave
functions, e.g. Fermion IC Molecular Dynamics
\cite{RevModPhys.72.655}, Electron Nuclear Dynamics
\cite{RevModPhys.66.917}, and time-dependent Multi configuration
Self-consistent-field calculations \cite{C7CP02086D}.
In these approaches, the wavefunction is
approximated by Slater determinants of localized single particle
orbitals. The orbitals are parameterized by dynamical variables 
(wave packet width, average position or momentum) and the TDVP is used 
to derive  equation of motion for these dynamical variables.
In this work we use the TDVP to time propagate a wave function by
optimizing its linear and nonlinear parameters. In a previous paper we used the imaginary time propagation method combined with the TDVP 
to accurately describe few-particle systems
\cite{PhysRevA.99.012504}. 
It was shown, that the TDVP can be used to
obtain basis functions with accuracy  comparable or
better than gradient based Newton-Raphson optimization. This success 
paves the way for the application of the TDVP to time-dependent problems.
We will test this application using the 1D and 3D Hydrogen atom with
Gaussian and  soft Coulomb potential in strong laser pulses, and then compare the results to finite difference 
grid calculations. The advantage of the present approach is that 
only a few basis functions
are needed, while in the finite difference calculations millions of
grid points are used. Moreover, the present approach can be extended to
larger systems, while the finite difference is limited to 3D. An
additional advantage is that no boundary conditions have to be
enforced, and the basis flexible evolves according to TDVP.

\section{Formalism}

\subsection{Time-dependent variational principle} 
The time dependent wave function of this system in a general form can
be written as:
\begin{equation}
\psi(t)=\psi({\bf q}(t))
\end{equation}
where ${\bf q}(t)$ is a set of linear and nonlinear variational
parameters.

The time-dependent Schr\"odinger equation,
\begin{equation}
i{d\over dt}\psi(t)=H\psi(t)
\end{equation}
will be solved by the McLachlan variational method
\cite{doi:10.1080/00268976400100041}. In this approach, the norm of the deviation between
the right-hand and the left-hand side of the time-dependent
Schr\"odinger equation is minimized with respect to the trial function.
The quantity 
\begin{equation}
  I = ||i \phi(t) -H \psi(t)||^2 \rightarrow \min
\label{tdv1}
\end{equation}
is to be varied with respect to $\phi$ only, and then the equivalency $\dot\psi\equiv \phi$ is enforced. At time $t$ the wave function is known and its time derivative is determined by
minimizing $I$. In case of $I=0$, an exact solution exists, but the
approximation in the expansion of $\psi(t)$ leads to $I>0$ values.

The variations of $I$ with respect to $\phi$ gives the equations of motion:
\begin{equation}
\lela \frac{\partial\psi}{\partial{\bf q}} \Big| i\dot\psi - H \psi
\rira = 0 \; .
\label{tdv2}
\end{equation}
%
This equation can be used to determine the (linear and nonlinear) variational parameters.

\subsection{Parameter optimization}

We can also write \eqref{tdv2} in matrix form as:
\begin{equation}
i M\dot {\bf q} = {\bf v} 
\label{prop}
\end{equation}
where 
\begin{equation}
M_{ij} = \lela\frac{\partial \psi}{\partial {q}_i}\Big|
\frac{\partial \psi}{\partial {q}_j}\rira \; , \; 	
\label{mmat}
\end{equation}
and
\begin{equation}
{v}_i = \lela \frac{\partial \psi}{\partial {q}_i}		 
\Big| H \psi \rira \; .
\end{equation}
		 
By approximating the time derivative with first order finite difference, 		 
Eq. \eqref{prop} becomes
\begin{equation}
\dot{\bf q} = -iM^{-1}{\bf v}
\label{tdv4}
\end{equation}
There are various established ways to solve such first order linear differential equations, and better approximations allowing larger time steps such as a Runge-Kutta approach can be used, but we elected to use the Euler method for time propagation for simplicity.

\subsection{Hamiltonian and basis functions}
We will test the approach by using a  Hamiltonian describing a particle in
a laser field in length  gauge
\begin{equation}
H=-{1\over 2}\left(
{d^2\over d x^2}+
{d^2\over d y^2}+
{d^2\over d z^2}\right)+V(x,y,z)+F(t)z,
\end{equation}
where $F(t)$ is the time dependent electric field pulse, which is defined as: 
\begin{equation}
F(t)=E_0 e^{-(t-T)^2/\tau^2}\cos(\omega t).
\label{laser}
\end{equation}
We define two different types of basis functions to represent the
time-dependent wave function. The first one takes on the form:
\begin{equation}
g_i=
c_i z^{n_i} g_{\alpha_i}(x)g_{\alpha_i}(y)g_{\beta_i}(z)
=c_i z^{n_i}  e^{-\alpha_i(x^2+y^2)-\beta_i z^2},
\label{basis1}
\end{equation}
where 
\begin{equation}
g_\sigma(x)=e^{-\sigma x^2}
\end{equation}
is a one dimensional Gaussian and will be referred to as polynomial times
Gaussian (PTG). The second basis is a plane wave times  Gaussian (PWG):
\begin{equation}
g_i=
c_i  g_{\alpha_i}(x)g_{\alpha_i}(y)g_{\beta_i}(z)e^{kz}
=c_i e^{-\alpha_i(x^2+y^2)-\beta_i z^2+kz}.
\label{basis2}
\end{equation}
The parameters of the Gaussians are kept equal in the $x$ and $y$
direction due to the cylindrical  symmetry of the potential. In one
dimensional (1D) test calculations $\alpha=0$ is used to reduce the
basis to 1D. 

The variational parameters form a vector,
\begin{equation}
{\bf q}(t)=\left(
\begin{array}{c}
c(t)\\
\alpha(t)\\
\beta(t)\\
\end{array}
\right)=
\left(
\begin{array}{c}
c_1(t)\\
{\vdots}\\
c_N(t)\\
\alpha_{1}(t)\\
{\vdots}\\
\alpha_{N}(t)\\
\beta_1(t)\\
{\vdots}\\
\beta_N(t)
\end{array}
\right),
\end{equation}
in the case of PTG and a similar vector can be defined for PWG.
For PTG, the values of $n_k$ must be set to be integers. The variational trial
function is
\begin{equation}
\psi(t)=\psi({\bf q}(t))=\sum_{k=1}^N c_k(t) \phi_k(t)=\sum_{k=1}^N g_k(t).
\label{exp}
\end{equation}

To illustrate the flexibility of the Gaussian basis in time-dependent
calculations, we solve the TDVP equation (Eq.
\eqref{prop}) analytically for a free particle in Appendix \ref{appA}. This case can be used
to test the time step and matrix elements in the numerical calculations.

As the example in Appendix \ref{appA} and Eq.\ \eqref{tdv4} show, the 
parameters of the basis functions become complex during the time
propagation. This is completely different from the conventional time
propagation in which the wave function is expanded into some basis, and the
linear coefficients are time dependent and complex.  A Gaussian with a
complex parameter can be written as:
\begin{equation}
e^{-(\alpha_r+i\alpha_i)x^2}=
e^{-\alpha_r x^2}\left(\cos(\alpha_i x^2)+i\sin(\alpha_i x^2)\right).
\end{equation}
This function is an oscillatory function with a Gaussian envelope, and
seems to greatly enhance the flexibility of the basis function 
\cite{PhysRevA.99.012504}. To make the integrals of the matrix element
convergent,  $\alpha_r$ should be positive, which is not explicitly guaranteed 
in the time propagation of Eq. \eqref{tdv4}, but in our numerical 
examples it was automatically satisfied.

We will use two potentials to test the approach. A single Gaussian
potential,
\begin{equation}
V=-V_0 e^{-\mu (x^2+y^2+z^2)},
\end{equation}
with $V_0=1$ and $\mu=0.1$ a.u., and 
a soft Coulomb potential,
\begin{equation}
V=-{1\over \sqrt{x^2+y^2+z^2+a^2}},
\end{equation}
with $a=1.0$ a.u., respectively.  We will use the soft Coulomb
potential because the Coulomb potential cannot be easily used in grid 
calculations and its use is problematic in 1D \cite{doi:10.1139/p06-072}.
In the case of the PWG basis, the soft Coulomb potential is expanded
into 50 Gaussians to facilitate the analytical calculations of the
matrix elements.

In 1D test cases, the condition $x=y=0$ is set in the potential and $\alpha=0$ is used 
in the basis function with 1D kinetic energy. The matrix elements of these 
basis functions can be calculated
analytically as it is shown in Appendices \ref{appb}, \ref{appc} and
\ref{appd}.

\subsection{Time propagation of the wave function}
Equation \eqref{tdv4} defines the time propagation of both linear and
nonlinear parameters of the wave function. With the exception of very
small time steps, the simple first order
finite difference approximation is not expected to be accurate enough
to preserve the norm of the wave function. To alleviate this problem, we only use this equation to time
propagate the nonlinear parameters and we update the linear parameters 
separately to preserve the norm. One can view this as an optimization 
of the basis functions by updating the nonlinear parameters using TDVP. 
We then time propagate the wave function on the updated basis. 

We have a set of basis function in time $t$, $\phi_k(t)$, which is
time-propagated to time $t+\Delta t$ to become $\phi_k(t+\Delta t)$
using Eq. \eqref{tdv4}. Both of these sets of basis functions 
can be used to represent the wave function at time $t$:
\begin{equation}
\psi(t)=\sum_{k=1}^N \hat{c}_k(t,t) \phi_k(t)=\sum_{k=1}^N
\hat{c}_k(t,t+\Delta t)
\phi_k(t+\Delta t).
\label{texp}
\end{equation}
In this equation $\hat{c}_k(t,t)$ is  known as we know the wave function at time $t$
(and it is not calculated using Eq. \eqref{tdv4}).
The unknown $\hat{c}_k(t,t+\Delta t)$ coefficients can be easily
derived by defining the overlap of the basis functions
\begin{equation}
S_{ij}(t,t')=\langle \phi_i(t)\vert\phi_j(t')\rangle
\end{equation}
and multiplying Eq. \eqref{texp} with $\psi_i(t)$. The result is:
\begin{equation}
\hat{c}_i(t,t+\Delta t)=\sum_{j=1}^N S^{-1}_{ij}(t,t)\sum_{k=1}^N
S_{j,k}(t,t+\Delta t) \hat{c}_k (t,t).
\end{equation}
Now we know the linear combination coefficient of the wave function
$\psi(t)$ at time $t$ on the optimal basis $\phi_k(t+\Delta)$, so we can
time propagate the wave function in the conventional way using
\begin{equation}
\psi(t+\Delta t)=e^{-iH\Delta t} \psi(t)
\end{equation}
to calculate $\hat{c}_k(t+\Delta t,t+\Delta t)$. We choose the
numerically stable Crank-Nicolson approach to update the coefficients:
\begin{eqnarray}
&&\hat{C}(t+\Delta t,t+\Delta t) = \\
&&{S(t+\Delta t,t+\Delta t)-{i\over 2}H(t+\Delta t,t+\Delta t)\over
 S(t+\Delta t,t+\Delta t)+{i\over 2}H(t+\Delta t,t+\Delta t)}
\hat{C}(t,t+\Delta t),
\nonumber
\end{eqnarray}
where $\hat{C}^T=(\hat{c}_1,{\ldots},\hat{c}_N)$ and 
\begin{equation}
H_{ij}(t,t')=\langle \psi_i(t)\vert H \vert \psi_j(t')\rangle.
\end{equation}
This approach significantly improves the stability of the approach and
allows larger time steps. 

\section{Calculations}

\subsection{Ground state}
Before the time propagation we need to calculate the ground state
(without the laser field). In the time propagation that will be the
initial state at $t=0$. To calculate the ground state the parameters 
of the Gaussians will be defined with a geometric progression,
\begin{equation}
{1\over \sqrt{\alpha_i}}=a \nu^{i-1},
\end{equation}
with $a=0.5$ and $\nu=1.3$.
For the ground state calculation, we will use $n=0$ for the PTG basis
and $k=0$ in the PWG basis.
For 1D grid calculation, $N=5000$ equidistant grid points are used with 
$h=0.125$ grid spacing, and a $N=61\times 61\times 1200$ size grid with
$h=0.25$ is used in 3D. While very fine grid spacing can be used in 1D, 
it must be larger in 3D due to the increase in computational cost.

The ground state energies are listed in Table I. {\rev These energies were
calculated by diagonalization of the PTG and PWG case. In the case of
the grid calculations, the ground state energy was calculated by the
conjugate gradient method using the codes of \cite{varga_book_cn}.} There is an excellent
agreement in 1D, and a slight difference between the PWG
and the grid calculation in 3D. While agreement can be achieved with a finer grid, 
there are more computational constraints the finer the grid becomes. We only used the PTG
for the Gauss potential, so the PTG ground state energy for other cases is not shown. 
\begin{table}[]
\begin{tabular}{|l|l|l|l|}
\hline
Basis   & $N$ & Potential        & Energy        \\\hline\hline
1D PTG  & 30        & Gauss            & -0.79526702   \\\hline
1D PWG  & 20        & Gauss            & -0.79526702   \\\hline
1D Grid & 5000      & Gauss            & -0.79526702   \\\hline
1D PWG  & 20        & Soft Coulomb     & -0.66977138   \\\hline
1D Grid & 5000      & Soft Coulomb     & -0.66977138   \\\hline
3D PWG  & 30        & Soft Coulomb     & -0.27489135   \\\hline
3D Grid & 4465200   & Soft Coulomb     & -0.27461231   \\\hline
\end{tabular}
\caption{Ground state energies (in a.u.). {\rev The basis dimension is $N$
for the PWG and PTG, and the number of grid points in the 1D and 3D
grid case.}}
\end{table}

\begin{figure}
\includegraphics[width=0.95\linewidth]{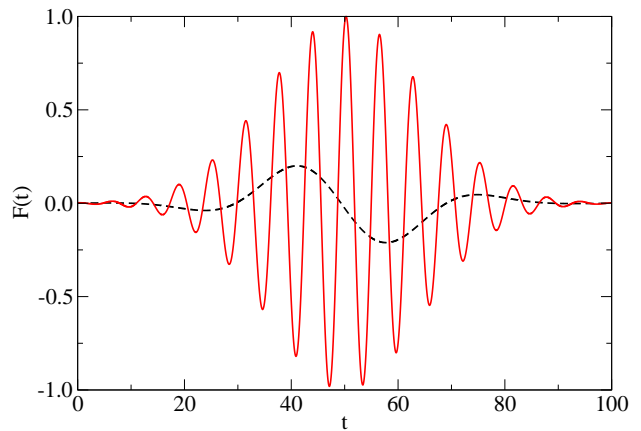} 
\caption{The laser fields used in the calculation, 
laser $A$, $E_0=0.25, \tau=20.5, \omega=1.0/2\pi, T=50$ (black dotted
line); laser $B$, $E_0=1.0, \tau=20.5, \omega=1.0, T=50$ (red line)}
\label{fig1}
\end{figure}

\subsection{Time propagation}
Two different laser pulses are used in the calculation. The first (see
Fig. \ref{fig1}), laser $A$, has only a few cycles and moves the electron 
to one direction as will be shown later. The second, laser $B$, has many cycles 
and moves the electron almost symmetrically left and right. The time 
step is $\Delta t=0.001$ a.u. in 1D calculations, and $\Delta
t=0.0005$ a.u. in the 3D calculations for both the PWG and the grid. The PTG requires a smaller time step as we will discuss later. 

The PTG ground state calculation was restricted to $n=0$ and
to make a starting PTG basis for time propagation, the basis will be doubled by adding $n=1$ states with the same $\beta_i$ parameters as
of the $n=0$ states. These states are needed because the laser field operator 
$F(t)z$ matrix elements are  only nonzero for basis states for even $n_i+n_{i'}+1$. To start
the calculation from the ground state, the linear coefficients of the
$n_i=1$ basis states will be set to zero. States with $n>1$ do not seem
to improve the calculation. The PWG basis does not need
any modification and one can start the computation form the ground state 
wave function.

\begin{figure}
\includegraphics[angle=270,origin=c,width=0.95\linewidth]{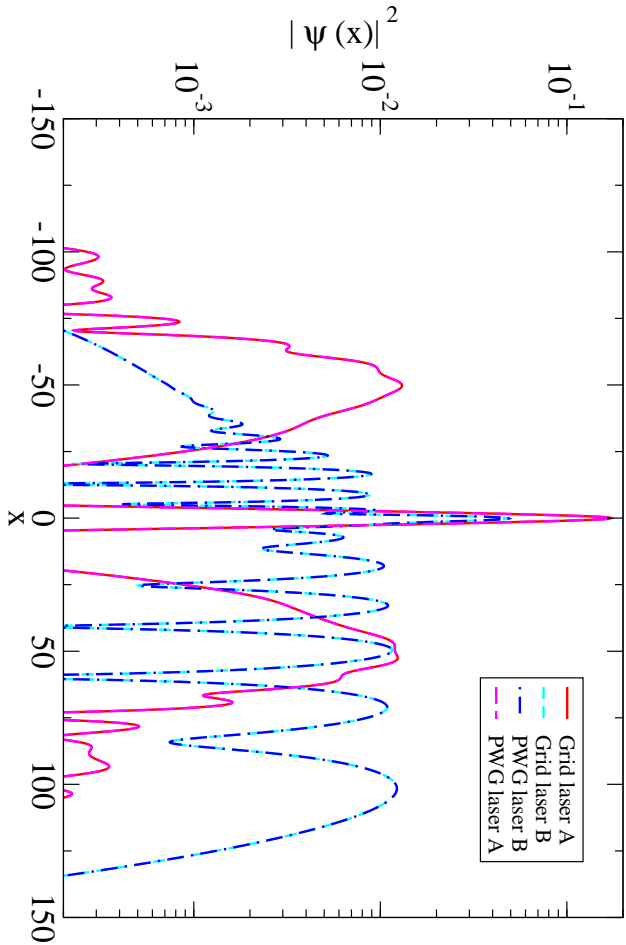} 
\caption{Electron densities in Gaussian potential at $t=100$ a.u. for
laser $A$  and laser $B$.}
\label{fig2}
\end{figure}

The electron density, $\vert\psi(x,t)\vert^2$, after time propagation
up to t=100 a.u. are compared in Fig. \ref{fig2} in the case of the Gaussian potential. 
The agreement between the grid and the PWG calculations are excellent.
In the asymptotic region where the density becomes smaller than
$10^{-4}$, the two approaches do not fully agree. This is partly because of numerical noise, which can be decreased with a smaller time step, and partly
due to the grid spacing. 

Test calculations show that PTG basis can only be used with smaller
time steps ($\Delta t=0.00001$ a.u.) to produce the same results as
the grid and PWG. This is because this basis easily becomes nearly 
linearly dependent (large overlap between basis functions),
especially in the 1D case, which makes the calculation of the inverse of $M$ difficult. The other difficulty is 
choosing the  optimal number of basis states with $n=0$ and $n=1$.
It is still useful to consider the PTG basis as an alternative test,
especially that in 3D the Coulomb potential can be analytically calculated for this basis (see Appendix \ref{appc}). 

Figures \ref{fig3}, and \ref{fig4} show the energy and the occupation probability of the ground state as a function of time. The occupation probability is defined as:
\begin{equation}
P(t)=\vert \langle \psi(0)\vert\psi(t)\rangle \vert^2.
\end{equation}
The energy and the occupational probability are in excellent
agreement for the grid, PTG, and PWG basis functions for both laser fields. Laser $A$ strongly ionizes the system and the ground
state occupation becomes about 0.3 after the pulse. This means 
(see Fig. \ref{fig2}) that the tail of the wave function has large
amplitude far away form the center of the potential, but the complex Gaussian basis is flexible enough to represent this. 

The next example is a test for a soft Coulomb potential. Since the PTG requires much smaller time step, we exclude it from the 
discussion from now. Figures
\ref{fig5} and \ref{fig6} show that the approach works well for the soft Coulomb potential as well. Comparing Figs. \ref{fig3} and \ref{fig4} to
\ref{fig5} and \ref{fig6} show that the
effect of the laser field is very similar in both the Gauss and soft 
Coulomb potentials. The electron is slightly less bound in the soft Coulomb potential and the laser causes larger excitation and ionization.
\begin{figure}
\includegraphics[width=0.95\linewidth]{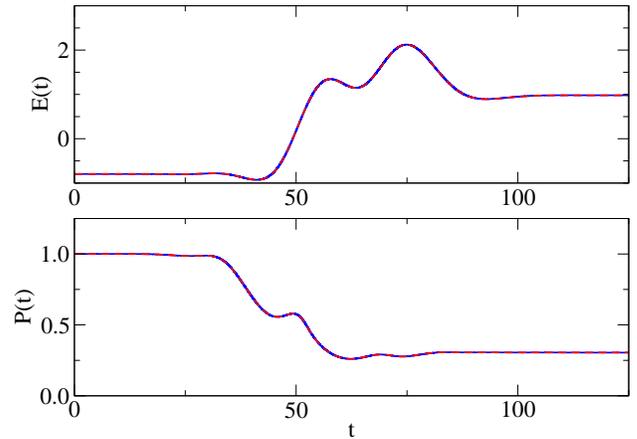} 
\caption{Gaussian potential with laser field $A$ in 1D. 
Top: Energy as a function of time for grid (solid blue line),
PWG (red dashed line) and PTG (black dotted line). 
Bottom: Ground state occupation probability  as a function of time for grid (solid blue line),
PWG (red dashed line) and PTG (black dotted line). The three lines are
indistinguishable in the resolution of the figure.}
\label{fig3}
\end{figure}

\begin{figure}
\includegraphics[width=0.95\linewidth]{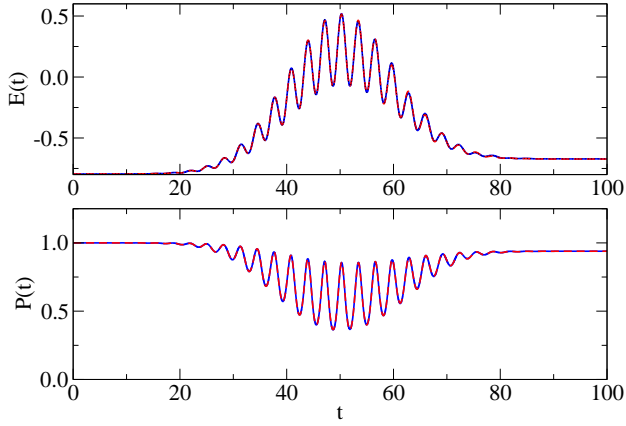} 
\caption{Gaussian potential with laser field $B$ in 1D. Top: Energy as a function of time for grid (solid blue line),
PWG (red dashed line) and PTG (black dotted line). 
Bottom: Ground state occupation probability  as a function of time for grid (solid blue line),
PWG (red dashed line) and PTG (black dotted line). The three lines are
indistinguishable in the resolution of the figure.}
\label{fig4}
\end{figure}

The last example covers the case of soft Coulomb in 3D for lasers $A$ and $B$, which are illustrated in Figs. \ref{fig7} and \ref{fig8}. The agreement between the grid and PWG calculations is still very good, although the necessary time step to reach accuracy is smaller for PWG than in 1D. The grid calculation would converge with a time step that is 10 times larger, but we used the same
time step for both grid and PWG for consistency. However, even with a larger time step, the grid calculation, is very computationally demanding due to its large grid size. Indeed, its computational time takes at least two orders of magnitude  longer than that of the PWG for the soft Coulomb potential. 

We have also tested a restricted PTG basis, constraining the Gaussian 
to be spherically symmetric by choosing $\alpha=\beta$ in Eq. \eqref{basis2}. 
Test calculations for shorter, weaker pulses show good agreement between this restricted basis and the grid calculations, but this basis is not
flexible enough for accurate calculations in the test examples presented in this work. Despite this, the result is still noteworthy because
it may lead to an extension of Gaussian atomic orbitals for weak fields.

\begin{figure}
\includegraphics[width=0.95\linewidth]{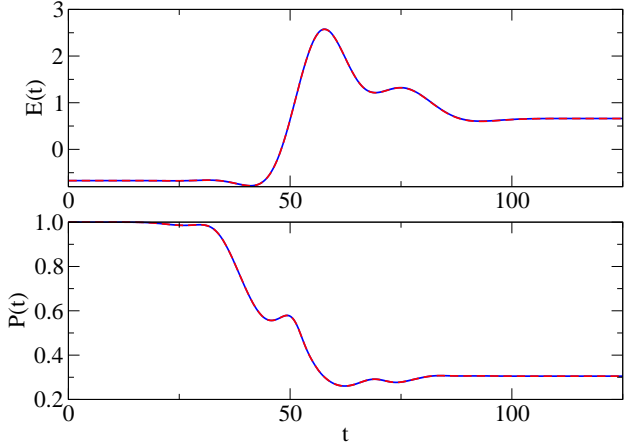} 
\caption{Soft Coulomb  potential with laser field $A$ in 1D.  
Top: Energy as a function of time for grid (solid blue line),
PWG (red dashed line). 
Bottom: Ground state occupation probability  as a function of time for grid (solid blue line),
PWG (red dashed line).}
\label{fig5}
\end{figure}

\begin{figure}
\includegraphics[width=0.95\linewidth]{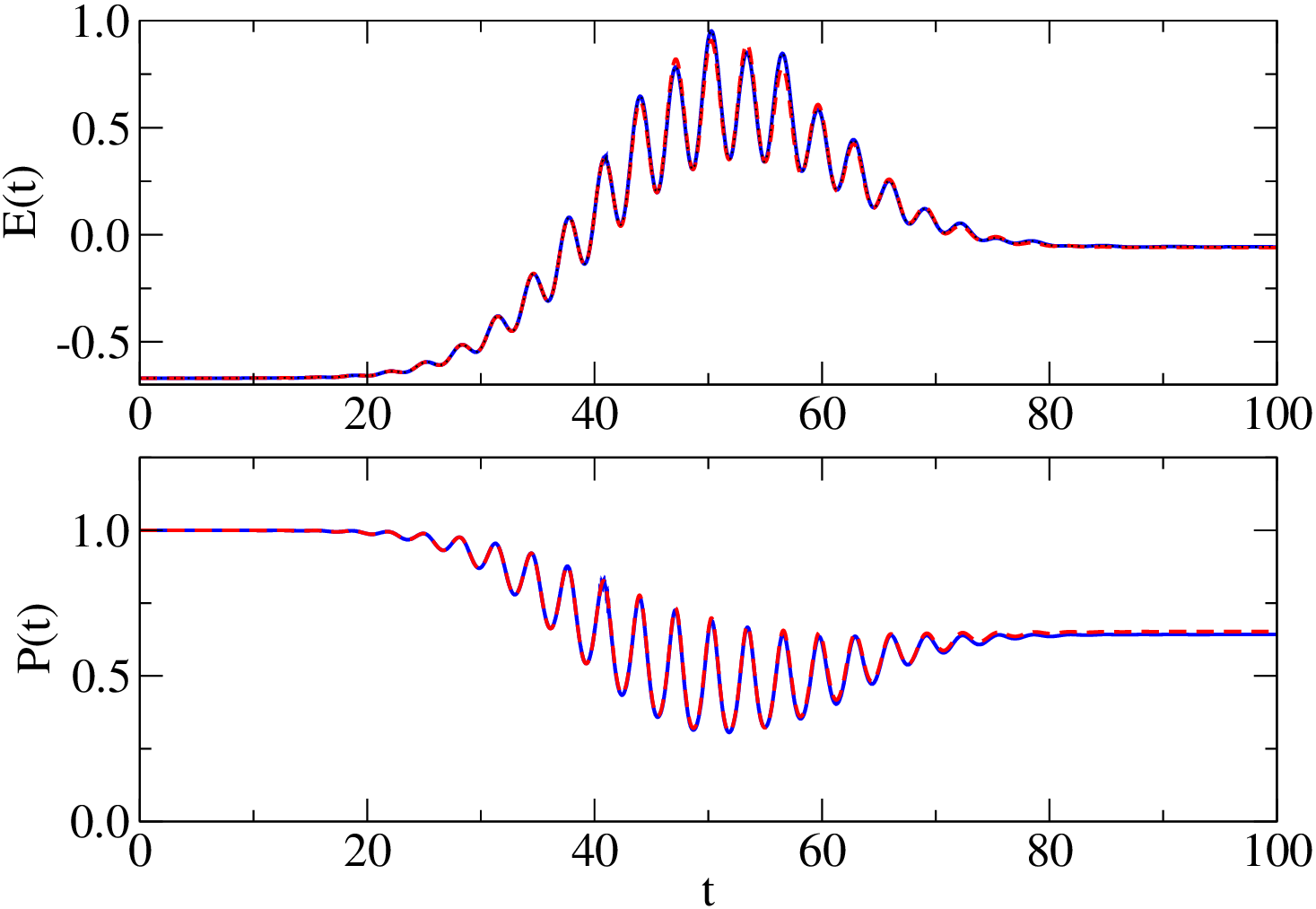} 
\caption{Soft Coulomb  potential with laser field $B$ in 1D.  
Top: Energy as a function of time for grid (solid blue line),
PWG (red dashed line). 
Bottom: Ground state occupation probability  as a function of time for grid (solid blue line),
PWG (red dashed line).}
\label{fig6}
\end{figure}

\begin{figure}
\includegraphics[width=0.95\linewidth]{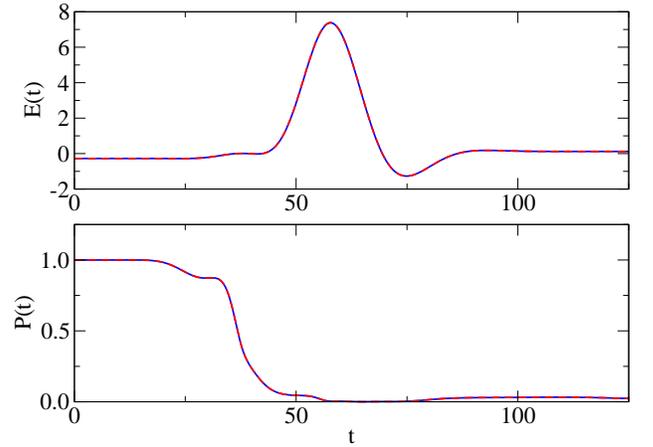} 
\caption{Soft Coulomb  potential with laser field $A$ in 3D.  
Top: Energy as a function of time for grid (solid blue line),
PWG (red dashed line). 
Bottom: Ground state occupation probability  as a function of time for grid (solid blue line),
PWG (red dashed line).}
\label{fig7}
\end{figure}

\begin{figure}
\includegraphics[width=0.95\linewidth]{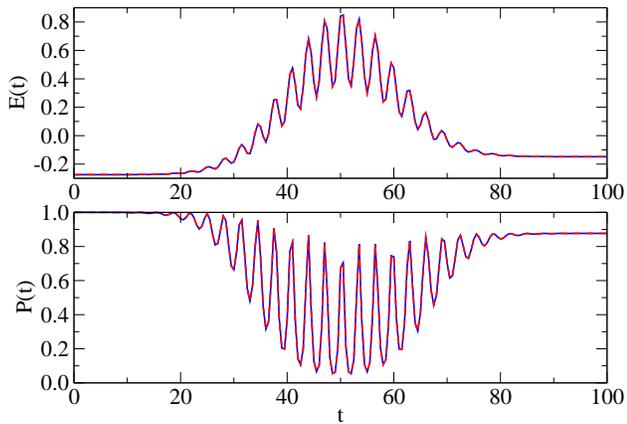} 
\caption{Soft Coulomb  potential with laser field $B$ in 3D.  
Top: Energy as a function of time for grid (solid blue line),
PWG (red dashed line). 
Bottom: Ground state occupation probability  as a function of time for grid (solid blue line),
PWG (red dashed line).}
\label{fig8}
\end{figure}

\rev{To test the applicability of the approach for larger systems we
have considered a two electron system in 1D with the Hamiltonian
\begin{equation}
H=-{1\over 2} {d^2\over dx_1^2} -{1\over 2} {d^2\over
dx_2^2}-2V(x_1)-2V(x_2)+V(x_1-x_2)+F(t)(x_1+x_2),
\end{equation}
with a Gaussian potential, $V(x)=e^{\mu x^2}$, $\mu=0.1$. The basis
function is taken in the form
\begin{equation}
g_i=c_ie^{-\alpha_{1i} x_{1}^2-\alpha_{2i} x_2^2+\beta_i x_1 x_2 +k_{1i}
x_1 +k_{2i} x_2}
\end{equation}
with six variational parameters,
$\alpha_{1i},\alpha_{2i},\beta_i,k_{1i},k_{2i}$ and $c_i$,
$(i=1,{\ldots},N$. The two particles are assumed to be distinguishable (one 
electron with spin up and one with spin down).

The energy of the two electron system as the function of time is shown
in Fig. \ref{fig9}. The convergence was checked by using different
starting basis sets and different basis dimensions. $N=15$ basis
functions with $\Delta t=0.0001$ a.u. yields well converged results. Figure
\ref{fig10} show the snapshots of the two-electron density. At $t=0$
the electrons are confined to the potential well around the origin.
The laser field moves them out of the well towards the positive
direction ($t=30$ a.u. in Fig. \ref{fig10}), and then back toward the origin. 
After the peak of the laser field (in Fig. \ref{fig10}, $t=50$ a.u.)
there are two peaks that appear in the density. This corresponds to a
configuration where the first electron's probability distribution has
a maximum close to the origin, while the second electron's probability
distribution has two maxima,
which are left and right with respect to the origin. 

\begin{figure}
\includegraphics[width=0.95\linewidth]{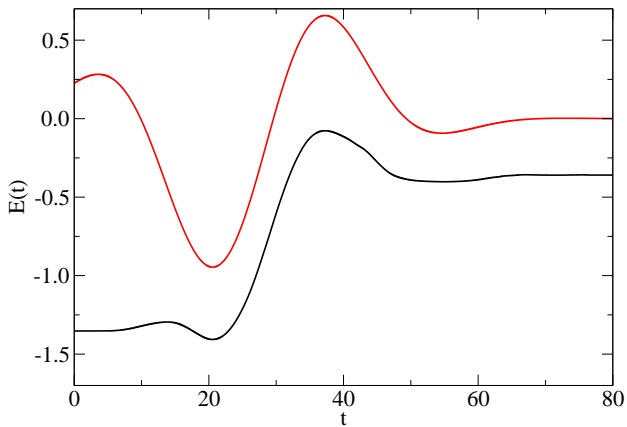} 
\caption{Energy (black line) and laser field (dashed line) of a 2
electron system as a function of time. The laser parameters are
$E_0=0.1, \tau=20.5, \omega=1.0, T=25$. The amplitude of the laser
on the figure is multiplied by 10 for better visibility.}
\label{fig9}
\end{figure}
\begin{figure}
\includegraphics[width=0.95\linewidth]{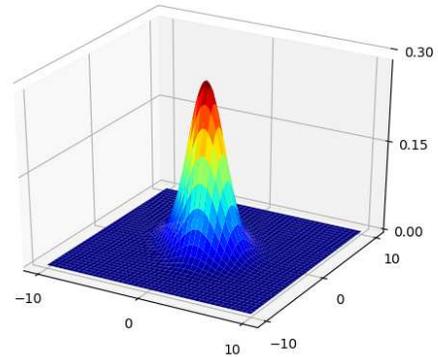} 
\includegraphics[width=0.95\linewidth]{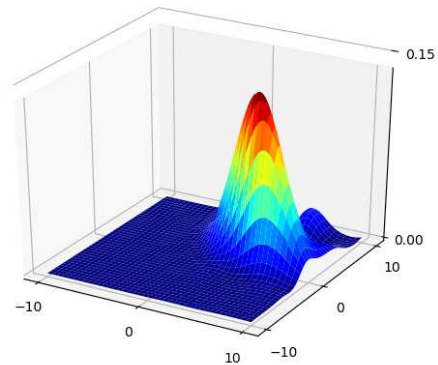} 
\includegraphics[width=0.95\linewidth]{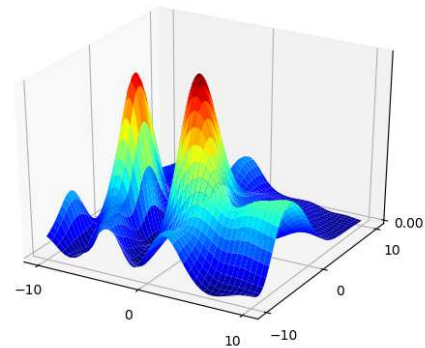} 
\caption{Snapshots of the two electron density at $t=0$, $t=30$ and
$t=50$ a.u.. The plane axis are $x_1$ and $x_2$, the coordinates of
electron 1 and 2.}
\label{fig10}
\end{figure}

}

\section{Summary}
We have used the TDVP to time propagate the wave function. The TDVP
optimizes the linear and the nonlinear
parameters on the same footing. The results are compared to those of grid calculations and the accuracy to the present approach is demonstrated. We have tested various forms of basis functions including Gaussians multiplied by polynomials, plane waves, and non-spherical Gaussians. 
The complex parameters of the Gaussians make the basis functions flexible enough to represent oscillatory wave functions. In addition, several
potentials and laser fields were used to test the approach for different degree of ionization.

The approach has several advantages. First, a simple Gaussian basis can be used to solve time-dependent problems, which may be useful in
various electronic structure codes. Second, the number of basis functions needed is considerably smaller than the number of grid points required to
represent a wave function, which makes the calculation faster. Furthermore, no boundary
conditions need to be enforced, and the TDVP automatically generates the Gaussians to represent the wave function in space. As the free
Gaussian wave packet example (Appendix \ref{appA}) shows, the wave function 
can propagate from any given point to any desired distance without artificial
reflections. In principle, a complex absorbing potential can also be used, in which case the number of Gaussian basis states may be less,
because the wave function only need to be represented in a well defined 
region. The approach can be extended to larger systems  using
Explicitly Correlated Gaussians \cite{RevModPhys.85.693}. \rev{The
example of a 2 electron system presented in this paper shows promising results.}

The main disadvantage is that the basis needs to be carefully initialized, otherwise 
large overlap between basis functions can make the inversion of the 
$M$ matrix in Eq. \eqref{mmat} difficult. 
This can possibly be alleviated by using a singular value decomposition for 
calculation of the inverse. It is also somewhat 
difficult to determine a sufficient number of basis functions and 
their desired initial paremeters to minimize error during time propagation.  

The approach can be improved in several ways. Chief among them, the simple first order time propagation 
should be replaced with a more accurate approach. The approach can also benefit 
from adaptive time steps, using larger time step for smooth regions of
the time dependent potential and smaller time steps where the
potential has abrupt changes. Both of these improvements would allow for larger time steps 
and faster calculation. One can also design some scheme
to prune the number of Gaussians and add new Gaussians as needed. Finally, another possibility is to refit the wave function
with a completely new set of Gaussians after a certain time interval to exclude ill-behaved basis states. 

\section{Acknowledgment}
This work has been supported by the National Science
Foundation (NSF) under Grant  No. IRES 1826917.

%


\appendix

\section{Propagation of a free Gaussian wave packet}
\label{appA}
To illustrate the time propagation on a simple example, we consider
the time evolution of a Gaussian wave packet in 1D. 
The Hamiltonian is:
\begin{equation}
H=-{1\over 2}
{\partial^2\over \partial x^2},
\end{equation}
and the trial function is written in the form
Defining a one dimensional Gaussian as:
\begin{equation}
g=e^{\gamma-\alpha x^2-\kappa x},
\end{equation}.
The derivatives with respect to the  parameters are:
\begin{equation}
{\partial g \over \partial\gamma}=g, \ \ \ 
{\partial g \over \partial\alpha}=-x^2g, \ \ \ 
{\partial g \over \partial\kappa}=-x g,
\end{equation}
so the $M$ matrix (see  Eq. \eqref{mmat}) is:
\begin{widetext}
\begin{equation}
M=\left(
\begin{array}{ccc}
\langle g\vert  g\rangle &
-\langle g\vert x^2 \vert g\rangle &
-\langle g\vert x \vert g\rangle \\
-\langle g\vert x^2 \vert g\rangle &
\langle g\vert x^4 \vert g\rangle &
\langle g\vert x^3 \vert g\rangle \\
-\langle g\vert x \vert g\rangle &
\langle g\vert x^3 \vert g\rangle &
\langle g\vert x^2 \vert g\rangle \\
\end{array}
\right).
\label{mmat1}
\end{equation}
\end{widetext}
The action of the Hamiltonian on the trial function can be expressed as:
\begin{equation}
Hg=-{1\over 2}{\partial^2 g\over \partial x^2}=
\left((\alpha-\kappa^2/2)-2\alpha^2 x^2-2\alpha\kappa x\right)g.
\label{ham}
\end{equation}
The ${\bf v}$ vector is defined as:
\begin{equation}
{\bf v}=\left(
\begin{array}{c}
\langle g\vert H\vert  g \rangle \\
-\langle g\vert x^2 H \vert g\rangle \\
-\langle g\vert x H \vert g \rangle \\
\end{array}
\right), 
\end{equation}
which can be rewritten using Eq. \eqref{ham} and the definition of $M$ as:
\begin{equation}
{\bf v}=M
\left(
\begin{array}{c}
\alpha-\kappa^2/2\\
2\alpha^2 \\
2\alpha \kappa \\
\end{array}
\right).
\end{equation}
Using this Eq. \eqref{prop} becomes:
\begin{equation}
iM
\left(
\begin{array}{c}
\dot \gamma\\
\dot \alpha \\
\dot \kappa \\
\end{array}
\right)
=
M\left(
\begin{array}{c}
\alpha-\kappa^2/2\\
2\alpha^2 \\
2\alpha \kappa \\
\end{array}
\right).
\end{equation}
The equation for $\alpha$,
\begin{equation}
i\dot \alpha=2\alpha^{2},
\end{equation}
can be integrated easily:
\begin{equation}
\alpha(t)={\alpha(0)\over 2i\alpha(0)t+1}.
\end{equation}
Substituting this into 
\begin{equation}
i\dot \kappa=2\alpha\kappa,
\end{equation}
we get
\begin{equation}
\kappa(t)= {\kappa (0) \over 2i\alpha(0)t+1}.
\end{equation}
Now using
\begin{equation}
    i\dot{\gamma} = \alpha - \kappa^2/2,
\end{equation}
we get
\begin{equation}
    \gamma(t) = -\frac{1}{2} \ln(2i\alpha(0) t +1) + \frac{i\kappa(0)^2t}{4i\alpha(0)t+2} + \gamma(0).
\end{equation}

The solution agrees with the analytical solution of time propagation of Gaussian wave packets.

\section{Matrix elements: 1D PTG}
\label{appb}
For simplicity, first we calculate the matrix elements for a single basis function:
\begin{equation}
g=z^{n} e^{\gamma-\beta z^2},
\end{equation}
and the we show how to generalize the results for $N$ basis functions. Instead of using using the linear coefficient $c$  we use $c=e^\gamma$, which makes the equations simpler: The derivative of the exponential function is proportional to the  exponential so the basis function remains in the same form.  In the ground state calculations, $c$ is a real
number, so to initialize $\gamma$ we set Im$(\gamma)=0$ if $c>0$ and
Im$(\gamma=\pi)$ if $c<0$. Alternatively, 
one can write the matrix elements in terms of $\gamma$ and switch back to $c$ in the numerical
work.

We take the derivatives with respect to the parameters:
\begin{equation}
    \frac{dg}{d\gamma} = g
\end{equation}

\begin{equation}
    \frac{dg}{d\beta} = -z^2g.
\end{equation}

We then get the matrix $M$ from Eq. \eqref{mmat}:

\begin{equation}
    M = \begin{pmatrix}
    \bra{g}\ket{g} & -\expval{z^2}{g} \\
    -\expval{z^2}{g} & \expval{z^4}{g}
    \end{pmatrix}.
    \label{ap2mat}
\end{equation}

Then $\textbf{v}$ is defined as:
\begin{equation}
    \textbf{v} = 
    \begin{pmatrix}
    \expval{H}{g} \\
    - \expval{z^2H}{g}
    \end{pmatrix}.
\end{equation}

To calculate these matrix elements we need the action of the kinetic energy operator on g:

\begin{equation}
-{1\over 2}{\partial^2 g\over \partial z^2}=
\left(-{1\over2}n(n-1)z^{-2}+(2n+1)\beta-2\beta^2z^{2}\right)g.
\end{equation}

The generalization for $N$ basis functions, $g_1,{\ldots} ,g_N$ is simple. The $M$ matrix in Eq. \eqref{ap2mat} will now be built up in $N\times N$ block matrices:

\begin{equation}
    M = \begin{pmatrix}
    \bra{g_i}\ket{g_j} & -\bra{g_i}{z^2}\ket{g_j} \\
    -\bra{g_i}{z^2}\ket{g_j} & \bra{g_i}{z^4}\ket{g_j}
    \end{pmatrix}.
\end{equation}

Similarly for \textbf{v} we have:
\begin{equation}
    \textbf{v} = \begin{pmatrix}
    \sum_{k=1}^N \bra{g_i} H \ket{g_k} \\
    -\sum_{k=1}^N \bra{g_i} z^2 H \ket{g_k}
    \end{pmatrix}.
\end{equation}

Now we assume a general potential can be expanded in terms of Gaussians:

\begin{equation}
    V(z) = \sum_i v_i e^{-v_z^i z^2}.
\end{equation}

In this case all the necessary matrix elements can be derived from:
\begin{equation}
\langle g_{\sigma}\vert z^{k}e^{-\nu z^2} \vert  g_{\sigma'} \rangle = e^{\gamma*+\gamma '}
{(k-1)!! \sqrt{\pi}\over (\sigma^*+\sigma'+\nu)^{(k+1)/2} 2^{(k/2)}}
\label{gme}
\end{equation}
if $k$ is even and zero otherwise. Note this formula is valid if the integral is convergent, 
which in turn is true if $\Re (\sigma^* + \sigma' + \nu) > 0$. The
principal value square root should be used in Eq. \eqref{gme}.

\section{Matrix elements: 3D PTG}
\label{appc}
In this section we calculate the matrix elements for a PTG basis function:
\begin{equation}
g=z^{n} e^{\gamma-\alpha(x^2+y^2)-\beta z^2}.
\end{equation}

We need the derivatives with respect to the  parameters:
\begin{equation}
{\partial g \over \partial\gamma}=g,
\end{equation}
\begin{equation}
{\partial g \over \partial\alpha}=-(x^2+y^2)g,
\end{equation}
\begin{equation}
{\partial g \over \partial\beta}=-z^2g.
\end{equation}
To calculate these matrix elements we need the action of the kinetic
energy operator on $g$:
\begin{equation}
-{1\over 2}\left({\partial^2\over \partial x^2}+{\partial^2\over
\partial y^2}\right)g=
\left(2\alpha-2\alpha^2(x^2+y^2)\right)g,
\end{equation}
\begin{equation}
-{1\over 2}{\partial^2 g\over \partial z^2}=
\left(-{1\over2}n(n-1)z^{-2}+(2n+1)\beta-2\beta^2z^{2}\right)g.
\end{equation}

The $M$ matrix in Eq. \eqref{mmat} will now be built up $N\times N$
block matrices:
\begin{widetext}
\begin{equation}
M=\left(
\begin{array}{ccc}
\langle g_i\vert  g_j\rangle &
-\langle g_i\vert x^2+y^2 \vert g_j\rangle &
-\langle g_i\vert z^2 \vert g_j\rangle \\
-\langle g_i\vert x^2+y^2 \vert g_j\rangle &
\langle g_i\vert (x^2+y^2)^2 \vert g_j\rangle &
\langle g_i\vert (x^2+y^2)z^2 \vert g_j\rangle \\
-\langle g_i\vert z^2 \vert g_j\rangle &
\langle g_i\vert (x^2+y^2)z^2 \vert g_j\rangle &
\langle g_i\vert z^4 \vert g_j\rangle \\
\end{array}
\right).
\end{equation}
\end{widetext}
Similarly, for the ${\bf v}$ vector we have: 
\begin{equation}
{\bf v}=\left(
\begin{array}{c}
\sum_{k=1}^N 
\langle g_i\vert H\vert  g_k \rangle \\
-\sum_{k=1}^N 
 \langle g_i\vert (x^2+y^2) H \vert g_k\rangle \\
-\sum_{k=1}^N 
\langle g_i\vert z^2 H \vert g_k \rangle \\
\end{array}
\right),
\end{equation}
where each entry corresponds to a $N \times 1$ block matrix.

Now we will assume, that a general potential can be expanded in terms
of Gaussians:
\begin{equation}
V(x,y,z)=\sum_{i} v_i 
e^{-\nu^i_x x^2-\nu^i_y y^2-\nu^i_z z^2}. 
\end{equation}
For spherically symmetric potentials this expansion further simplifies:
\begin{equation}
V(r)=\sum_{i} v_i e^{-\nu^i (x^2+y^2+z^2)}. 
\end{equation}
In  case of Gaussian potentials, all the necessary matrix elements of $M$ and
${\bf v}$ can be derived from:
\begin{equation}
\langle g_{\sigma_x}\vert x^{k_x}e^{-\nu_x x^2} \vert  g_{\sigma_x'} \rangle 
\langle g_{\sigma_y}\vert y^{k_y}e^{-\nu_y y^2} \vert  g_{\sigma_y'} \rangle 
\langle g_{\sigma_z}\vert z^{k_z}e^{-\nu_z z^2} \vert  g_{\sigma_z'} \rangle 
\label{me}
\end{equation}

The one dimensional integral can be easily calculated as above in Eq.
\eqref{gme}.

One can also calculate the matrix elements analytically for the 3D Coulomb
potential:
\begin{equation}
    V(r) = -{1\over{\sqrt{x^2+y^2+z^2}}}
\end{equation}

We can calculate the necessary matrix elements using the following integral identity:
\begin{equation}
    {1\over \sqrt{x^2+y^2+z^2}}= {2\over{\sqrt{\pi}}}\int_{0}^{\infty}{e^{-u^2(x^2+y^2+z^2)}}du.
\end{equation}

This allows us to evaluate the integral:
\begin{widetext}
\begin{equation}
\label{potential integral}
    \bra{g}V(x,y,z)\ket{g'}=  -{2\over{\sqrt{\pi}}}{e^{\gamma^* + \gamma'}}\int_{0}^{\infty}du\int_{-\infty}^{\infty}\int_{-\infty}^{\infty}\int_{-\infty}^{\infty} z^{n+n'}e^{-(u^2+a)(x^2+y^2)-(u^2+b)z^2}dxdydz,
\end{equation}
\end{widetext}
where 
\begin{equation}
    a = \alpha^* + \alpha',
    b = \beta^* + \beta'.
\end{equation}
With $n+n'=0$ \eqref{potential integral} yields:
\begin{equation}
    \label{no_poly_potential}
    \bra{g}V(x,y,z)\ket{g'}= -2\pi{e^{\gamma^* + \gamma'}}\frac{\arccos(\frac{\sqrt{a}}{\sqrt{b}})}{{\sqrt{a}\sqrt{b}\sqrt{1-\frac{a}{b}}}}
    ,
\end{equation}
where Re$(a)>0$ and Re$(b)>0$. In the case that $a=b$:

\begin{equation}
    \bra{g}V(x,y,z)\ket{g'}= -\frac{2\pi}{a}{e^{\gamma^* + \gamma'}}.
\end{equation}

Note that \eqref{potential integral} will be 0 when the polynomial terms are of odd degree. We can thus evaluate the more general form by taking derivatives as follows:

\begin{equation}
    (-1)^{n}\frac{\partial^n}{\partial a^n}\bra{g}V(x,y,z)\ket{g'}=\bra{g}(x^2+y^2)^nV(x,y,z)\ket{g'},
\end{equation}

\begin{equation}
    (-1)^{n}\frac{\partial^n}{\partial
    b^n}\bra{g}V(x,y,z)\ket{g'}=\bra{g}z^{2n}V(x,y,z)\ket{g'}
    ,
\end{equation}

\begin{equation}
    \frac{\partial^{2n}}{(\partial a \partial
    b)^n}\bra{g}V(x,y,z)\ket{g'}=\bra{g}(x^2+y^2)^{n}z^{2n}V(x,y,z)\ket{g'}
    .
\end{equation}

When $a=b$, the form \eqref{no_poly_potential} cannot be evaluated. 
In this case, \eqref{potential integral} simplifies considerably into a form which can be easily evaluated and yields simple polynomial
answers \cite{svm_book}.

\section{Matrix elements: 3D PWG}
\label{appd}
In this case we have the basis function in the form:
\begin{equation}
g=e^{\gamma-\alpha(x^2+y^2)-\beta z^2+kz}.
\end{equation}

The $M$ matrix in Eq. \eqref{mmat} will now be built up $N\times N$
block matrices:
\begin{widetext}
\begin{equation}
M=\left(
\begin{array}{cccc}
\langle g_i\vert  g_j\rangle &
-\langle g_i\vert x^2+y^2 \vert g_j\rangle &
-\langle g_i\vert z^2 \vert g_j\rangle & 
\langle g_i\vert z \vert  g_j\rangle 
\\
-\langle g_i\vert x^2+y^2 \vert g_j\rangle &
\langle g_i\vert (x^2+y^2)^2 \vert g_j\rangle &
\langle g_i\vert (x^2+y^2)z^2 \vert g_j\rangle &
-\langle g_i\vert (x^2+y^2) z \vert  g_j\rangle 
\\
-\langle g_i\vert z^2 \vert g_j\rangle &
\langle g_i\vert (x^2+y^2)z^2 \vert g_j\rangle &
\langle g_i\vert z^4 \vert g_j\rangle &
-\langle g_i\vert z^3 \vert  g_j\rangle  \\
\langle g_i\vert z \vert  g_j\rangle &
-\langle g_i\vert (x^2+y^2) z \vert  g_j\rangle &
-\langle g_i\vert z^3 \vert  g_j\rangle  &
\langle g_i\vert z^2 \vert  g_j\rangle  \\
\end{array}
\right).
\end{equation}
\end{widetext}
Similarly, for the ${\bf v}$ vector we have: 
\begin{equation}
{\bf v}=\left(
\begin{array}{c}
\sum_{k=1}^N 
\langle g_i\vert H\vert  g_k \rangle \\
-\sum_{k=1}^N 
 \langle g_i\vert (x^2+y^2) H \vert g_k\rangle \\
-\sum_{k=1}^N 
\langle g_i\vert z^2 H \vert g_k \rangle \\
\sum_{k=1}^N 
\langle g_i\vert z H \vert g_k \rangle \\
\end{array}
\right).
\end{equation}
All the necessary matrix elements can then be calculated from Eq.
\eqref{me} using Eq. \eqref{gme}  provided that the potential 
is expanded into Gaussians. The matrix elements for 1D PWG
can be obtained by taking $\alpha=0$ and  eliminating the second row and column from $M$,
and the second row from ${\bf v}$.

\end{document}